\documentclass[seceq]{ptptex}

\usepackage{graphicx}

\newcommand{\pimass}{m_{\pi}}

\newcommand{\la}{\langle}
\newcommand{\ra}{\rangle}

\title{%        %You can use \\ for explicit line-break
In-medium pions and partial restoration of chiral symmetry: 
\\
a model-independent analysis %
}

\author{%       %Use \scshape  for the family name
Daisuke \textsc{Jido}$^{1}$,
Tetsuo \textsc{Hatsuda}$^{2}$ and
Teiji \textsc{Kunihiro}$^{1}$ 
}

\inst{%         %Affiliation, neglected when [addenda] or [errata]
$^1$ Yukawa Institute for Theoretical Physics,
Kyoto University, Kyoto 606-8502, Japan\\
$^{2}$ Department of Physics, University of Tokyo, Tokyo, 113-0033, Japan
}

\abst{%         %this abstract is neglected when [addenda] or [errata]
Exploiting operator relations in QCD, we derive a 
novel and model-independent formula relating  
 the in-medium quark condensate $\la \bar{q}q\ra ^{*}$ to 
 the decay constant $F_t^*$ and the wave function renormalization
  constant $Z^*$ of the pion in the nuclear medium. 
 Evaluating  $Z^{*}$ at low density from  the iso-scalar
  pion-nucleon scattering data, it is concluded that 
  the enhanced repulsion of the $s$-wave isovector pion-nucleus
 interaction observed in the deeply bound pionic atoms  
 implies directly the reduction of the in-medium quark condensate.
 The knowlege of the in-medium pion mass is not necessary to
 reach this conclusion.
}

\begin{document}

\maketitle

\section{Introduction}

Exploring possible evidence of partial restoration of chiral symmetry in 
nuclear medium is one of the hot topics in modern hadron physics.
Systematic studies of deeply bound pionic atoms\cite{Suzuki:2002ae} 
and the low-energy $\pi^{-}$-nucleus elastic scattering\cite{Friedman:2004jh}
have shown an enhancement of the repulsion in 
the $s$-wave isovector pion-nucleus interaction. 
 This enhanced repulsion can be nicely
 accounted for by a reduction of the pion decay constant in nuclei 
provided that the $s$-wave isovector pion-nucleus interaction is given 
by the Weinberg-Tomozawa term with 
the {\em in-medium} pion decay constant $F_t^*$ \cite{Weise:2000xp}. 
The reduction of $F_t^*$ in nuclear medium is also
 argued to be the  mechanism of  the enhanced attraction
 of the in-medium $\pi\pi$ scattering in the $I=J=0$ channel (the $\sigma$ channel)
\cite{Hatsuda:1999kd}. This  has intimate relation to the 
near-threshold enhancement of the two-pion production off nuclei\cite{pipiexp}.

Recent theoretical works of chiral effective theory 
\cite{Jido:2000bw} and of the pion optical potential \cite{Kolomeitsev:2002gc}
have suggested that  the  in-medium renormalization of
 the pion wave function $Z^*$ is responsible for the 
  reduction of $F_t^*$.  The purpose of this work is to
present a new scaling law relating $F_t^*$ to the in-medium quark condensate 
$\la \bar{q}q\ra ^{*}$ and $Z^*$ \cite{JHK}. 
This relation is derived in a model-independent way 
based only on low energy 
theorems of QCD.
 Combining this relation and empirical observation
 of the pionic atoms and the $\pi$-nucleon scattering, 
 we conclude the reduction of the quark condensate 
in the nuclear medium without recourse to the knowledge
of the  in-medium pion mass.

\section{Derivation of the scaling law}
Let us consider the  correlation functions of the axial current 
$A_{\mu}\equiv \bar \psi \gamma_{\mu}\gamma_{5} t^{a} \psi$ 
and/or the pseudoscalar density 
$\phi_{5}^{a} \equiv \bar \psi i\gamma_{5} t^{a} \psi$, 
where $t^{a}$ is $1/2$ times the Pauli matrix.
 To discuss how the in-medium quark condensate is expressed 
by hadronic quantities, we start with the following correlation function 
for  the  symmetric nuclear matter in the chiral limit:
\begin{equation}
\Pi^{ab}(q) = \int d^4 x e^{iq\cdot x} \partial^\mu \langle \Omega | 
T[A_\mu^a(x) \phi_5^b(0)] | \Omega \rangle \label{eq:Aphicorr}
\end{equation}
where $|\Omega\rangle$ denotes a nuclear matter state normalized as 
$\langle \Omega | \Omega \rangle =1$.
Exploiting an operator relation, 
$
\partial^{\mu} TA_{\mu}^{a}(x) \phi_{5}^{b}(0)
 = T\partial^{\mu} A_{\mu}^{a}(x) \phi_{5}^{b}(0) + \delta(x_{0}) [A_{0}^{a}(x), \phi_{5}^{b}(0)]
$
with the conservation of the axial current, 
one can express the correlation function~(\ref{eq:Aphicorr}) 
by the quark condensate: 
\begin{eqnarray}
 \lim_{q_{\mu} \rightarrow 0} \Pi^{ab}(q_{\mu}) &=& 
     \int d^{4}x\,  \delta(x_{0}) \langle \Omega | 
  [A_{0}^{a}(x), \phi_{5}^{b}(0)] | \Omega \rangle 
 = - i \delta^{ab} \langle \Omega | \bar q q | \Omega \rangle  \ . \label{eq:FD2nd}
\end{eqnarray}
where we have used the commutation relation 
$[Q_{5}^{a}, \phi_{5}^{b}] = - i \delta^{ab} \phi$ with the generator 
$Q_{5}^{a}(x_{0}) = \int d^{3}x A_{0}^{a}(x_{0},x)$ and the scalar density 
$\phi=\bar \psi (I/2) \psi$.
In such a soft limit, the correlation function~(\ref{eq:Aphicorr}) is
saturated by zero modes whose contents in the chiral limit
 are the pionic mode and particle-hole excitations in nuclear matter.
These modes are coupled and mixed  with each other in nuclear matter.
 Here we are interested in the zero mode which is continuously
connected to the  pion state in the vacuum. 
Let us define $| \Omega_{5}^{a}\rangle$ as
such a pionic zero-mode state having pionic quantum number
 ($I=1$, $J^{P}=0^{-}$) and a momentum $q=(q_{0},\vec q)$. 
It can be shown that this state is continuously connected to 
the in-vacuum pion state \cite{JHK}. 
With the four vector $n_{\mu}$ characterizing the nuclear matter,
the conservation of the axial current and Lorentz invariance 
 lead to the following relations  for the matrix elements of 
$A_{\mu}^{a}$ and $\phi_{5}^{a}$ taken by the states $|\Omega \rangle$ 
and $| \Omega_{5}^{a} \rangle$: 
\begin{align}
  \langle \Omega | \phi^{a}_{5}(x) | \Omega^{b}_{5}(q) \rangle &= \delta^{ab} Z^{*1/2}e^{-iq\cdot x} , \label{eq:medZ}  \\ 
  \langle \Omega | A_{\mu}^{a}(x) | \Omega^{b}_{5}(q) \rangle & =
     \delta^{ab} i \left(- \frac{q^{2}}{(q\cdot n)} n_{\mu} +  q_{\mu}\right)  F^{*} e^{-iq\cdot x} . \label{eq:mateleA}
\end{align}
The matrix elements $Z^{*}$ and $F^{*}$ are  functions of $q\cdot n$ and $q^{2}$.
Taking the frame $n_{\mu}=(1,0,0,0)$ and 
using the in-medium dispersion relation $q_{0}^{\, 2}-v_{\pi}^{2} \vec q^{\: 2} = 0$ with the velocity $v_{\pi}$, 
we obtain linear dependence of the matrix elements on energy as follows: 
\begin{align}
   \langle \Omega | A_{0}^{a}(x) | \Omega_{5}^{b}(q) \rangle &=     \delta^{ab}
    i q_{0}\, \frac{1}{v_{\pi}^{2}} F^{*} e^{-iq\cdot x} ,
     \label{eq:medF} 
 &  \langle \Omega | A_{i}^{a}(x) | \Omega_{5}^{b}(q) \rangle =    \delta^{ab}
   i q_{i}\,  F^{*} e^{-iq\cdot x}   . % \label{eq:medFs}
\end{align}
Writing the temporal and spacial components of the decay constant 
separately as $F_{t}^{*} = F^{*}/v_{\pi}^{2}$ and $F_{s}^{*} = F^{*}$, 
one can see that the relation $F_{s}^{*}/F_{t}^{*} = v_{\pi}^{2}$ 
is automatically satisfied in this derivation \cite{JHK}.
Hereafter we refer to $F^{*}_{t,s}$ and $Z^{*}$ as the 
 quantities in the soft limit $q_{\mu}=0$. 
 The pion contribution to  the correlator Eq.(\ref{eq:Aphicorr})
can be  isolated in the soft limit and is rewritten as
\begin{equation}
  \lim_{q_{\mu}\rightarrow 0} \Pi^{ab}(q) = 
 \lim_{q_{\mu}\rightarrow 0} \left(i \delta^{ab}\frac{q_{0}^{2}F_{t}^{*}-\vec q^{\, 2}F_{s}^{*}} {q_{0}^{2}-v_{\pi}^{2} \vec q^{\, 2}}  Z^{*1/2} + \cdots \right)
  =  \delta^{ab} i F^{*}_{t}  Z^{*1/2} \ . \label{eq:FD1st}
\end{equation}
The ellipsis in Eq.\eqref{eq:FD1st} denotes terms without 
singularities at $q_{\mu} = 0 $, which are strongly suppressed 
in the soft limit due to the factor $q_{\mu}$ in the numerator. 
Combining Eqs.\ (\ref{eq:FD2nd}) and~(\ref{eq:FD1st}), 
we finally arrive at   an exact relation in the chiral limit 
\begin{equation}
  F_{t}^{*} Z^{*1/2} = -  \langle \Omega |  \bar q q | \Omega \rangle  \ .
  \label{eq:mFZqq}
\end{equation}
Taking a ratio with the relation in the vacuum, 
we find the following novel scaling law
connecting the pion decay constants and the quark 
condensate through the pion wave function renormalization constant:
\begin{equation}
   \frac{ F_t^* }{ F} \frac{Z^{*1/2}}{Z^{1/2}} = 
\frac{\langle \Omega | \bar q q | \Omega \rangle}{\langle 0| \bar qq | 0 \rangle}
\ .
\end{equation}
This relation  implies that one can
 deduce the in-medium reduction of the quark condensate 
solely from the decrease of the pion decay constant, 
once one knows the wave function renormalization of the pion in medium.

The in-medium Weinberg-Tomozawa relation can be also derived 
in a similar way as 
Eq.~\eqref{eq:mFZqq}. 
The essential point  is that 
the renormalization of the pion wave function for the isoscalar part is 
 already taken into account  and the isovector part is treated 
as a perturbation \cite{JHK}. 
The in-medium Gell-Mann--Oakes--Renner relation is 
also derived in this line with the PCAC relation under the assumptions 
of the small quark mass and the linear density approximation \cite{JHK}.

%

%%%%%%%%%%%%%%%%%%%%%%%%%%%%%%%%%%%%%%%%
%%%%%%%%%%%%%%%%%%%%%%%%%%%%%%%%%%%%%%%%

\section{Pion wave function renormalization in nuclear medium}

For examining the renormalization of the pion wave function 
in the low density limit,
let us consider the following correlation function of the pseudoscalar
density in symmetric nuclear matter in the chiral limit:
%\begin{equation}
$
  D_{\pi}^{*}(q) = 
  \int d^{4} x\, e^{i q \cdot x} \langle \Omega | T[ \phi_{5}(x)
  \phi_{5}(0)]
 | \Omega \rangle
$,
%  \ . \label{eq:medPcor} 
%\end{equation}
which  has the pion pole at 
$q^{2}_{0} - v_{\pi}^{2} \vec q^{\, 2} = 0$ with the residue $Z^{*}$ 
defined in Eq.\ (\ref{eq:medZ}). 
Collecting all the in-medium corrections
to the self-energy $\Sigma_{\pi}(q_{0},\vec q)$, or the optical
potential,
we can express the in-medium normalization constant by the self-energy as 
\begin{equation}
     Z^{*} = Z \left( 1 - \left. \frac{\partial \Sigma_{\pi}(q_{0},\vec q=0)}{\partial q_{0}^{2}} \right|_{q_{0}=0} \right)^{-1}  \ .  \label{eq:wfren}
\end{equation}
In the linear density approximation, 
the self-energy in symmetric nuclear matter is given by 
$\Sigma_{\pi}(q_{0}) = - \rho T_{\pi N}^{(+)} (q_{0})$ 
with an iso-singlet $\pi N$ scattering amplitude:
% given by the reduction formula:
%\begin{eqnarray}
$
%\lefteqn{ 
 T_{\pi N}^{(+)}(\nu,\tilde \nu;k^{2},k^{\prime2}) =
% } &&
%\nonumber \\ &&
 \, i Z^{-1}k^{2}k^{\prime2}
 \int d^{4}x \, e^{ik\cdot x} \langle N | T \phi_{5}(x) \phi_{5}(0) | N \rangle 
$
%\end{eqnarray}
with the outgoing (incoming) pion momentum $k$ ($k^{\prime}$), 
and kinematical variables defined 
by $\nu \equiv p_{N}\cdot (k+k^{\prime})/ (2 M_{N})$ 
and $ \tilde \nu  \equiv  - k^{\prime}\cdot k$. 
The off-shell extrapolation of this amplitude is defined 
by the above reduction formula and  
is consistent with the low energy theorems obtained 
%by current algebra s to chiral symmetry of the pseudoscalar
from the commutation relation involving the pseudoscalar
density $\phi_{5}$. The contributions to the off-shell 
amplitudes are represented by the diagrams in which the pseudoscalar 
density couples directly to the nucleon states~\cite{Thorsson:1995rj}.
The $\pi N$ scattering amplitude with $\vec q = 0$ is a function of 
$q_{0}$; $\nu^{2},\tilde \nu, k^{2}, k^{\prime 2}\rightarrow q_{0}^{2}$.
The chiral expansion is valid for the low energies and we have
\begin{equation}
   T_{\pi N}^{(+)}(q_{0}) = \alpha + \beta q_{0}^{2}  
\end{equation}
up to the higher order terms.
The intercept $\alpha$ is given by the explicit chiral symmetry 
breaking proportional to the quark mass, 
while the slope $\beta$ is a value in the chiral limit. 
From Eq.(\ref{eq:wfren}), one sees that 
$Z^{*}$ is given by the slope $\beta$ in the low density limit.
The sign of $\beta$ can be extracted as follows:
(i) In the Weinberg point, $\nu,\tilde \nu, k^{2}, k^{\prime 2} = 0$, 
we have $T_{\pi N}^{(+)}(0) = -\sigma_{\pi N}/F^{2} = \alpha$ 
with the $\pi N$ sigma term 
$\sigma_{\pi N}$~\cite{Weinberg:1966kf,Cheng:1970mx}. (ii)  At the
threshold, the amplitude is identified as the scattering length~$a_{\pi N}$, 
$T_{\pi N}^{(+)}(\pimass)  = 4\pi (1+\pimass/m_{N}) a_{\pi N}  =
\alpha + \beta \pimass^{2}$. 
 The pionic hydrogen atom data suggest a very small scattering length 
$a_{\pi N}=(0.0016\pm 0.0013)\pimass^{-1}$~\cite{Schroder:1999uq}, 
which is consistent with zero. Combing these facts, we find 
$\beta \simeq \sigma_{\pi N}/(F^{2}\pimass^{2})$.
Since the value of the $\pi N$ sigma term is positive, 
we can conclude that
\begin{equation}
  Z^{*1/2}/ Z^{1/2}\simeq F^2\pimass^2/(F^2\pimass^2+\rho \sigma_{\pi N}) < 1 
 % Z^{*1/2}/ Z^{1/2}\simeq 1/(1+\rho \sigma_{\pi N}/F^2\pimass^2) < 1 
  \label{eq:redZ}
\end{equation}
This conclusion is not altered as long as the scattering length is larger 
than $-0.050 \pimass^{-1}$  for $\sigma_{\pi N} \simeq 45$ MeV 
and  $-0.067 \pimass^{-1}$ for $\sigma_{\pi N} \simeq 60$ MeV.

\section{Summary}

On the basis of exact low-energy theorems in QCD, 
 we have derived a new scaling law which relates the in-medium quark
condensate to the in-medium pion decay constant 
 and  the in-medium pion wave function 
renormalization constant.   
We have utilized operator relations and chiral symmetry
  in QCD so that an explicit analysis of 
complicated dynamics of the pion in nuclear matter
is partially circumvented. 
Combining the new relation with experimental data of the pionic atoms
and $\pi N$ scattering, 
we conclude the reduction of the quark condensate in nuclear medium. 

\section*{Acknowledgements}
D.J.\ thanks Professor W. Weise and Professor M.\ Harada 
for fruitful discussions on this work during YKIS2006.
This work was supported in part by Grant-in-aid for Scientific Research
of Monbukagakusho of Japan (Nos.17540250, 18042001, 18540253) 
and by the 21st Century COE ``Center for
Diversity and Universality in Physics"  of Kyoto University.
%from the Ministry of Education, Culture, Sports, 
%Science and Technology of Japan.

%\bibliographystyle{nuclphys.bst}
%\bibliography{references}% Produces the bibliography via BibTeX.

\end{document}